\documentclass[twoside]{article}
\usepackage[a4paper]{geometry}
\usepackage{RR}

\usepackage{amssymb}
\usepackage{amsmath}
\usepackage{graphicx}
\usepackage{leftidx}
\usepackage[noend]{algpseudocode}
\usepackage{algorithm}
\usepackage{hyperref}

\newcommand{\R}{\mathbb{R}}
\newcommand{\N}{\mathbb{N}}
\newcommand{\IR}{\mathbb{IR}}

\newcommand {\mBox}[1] {\mathbf{#1}}
\newcommand {\mInt}[1] {\mathbf{#1}}
\newcommand {\intExt}[1] {\square {#1}}
\newcommand {\interior}[1] {int(#1)}
\newcommand {\border}[1] {\partial(#1)}
\newcommand {\borneinf}[1] {l({#1})}
\newcommand {\bornesup}[1] {u({#1})}
\newcommand {\width}[1] {w({#1})}
\newcommand {\middl}[1] {m({#1})}

\newcommand {\floatSet}[1] {\mathop{}\mathopen{\vphantom{#1}}^{[#1]}\kern-\scriptspace \R}

\newcommand {\EvalOrderZero}[1] { \mathop{}\mathopen{\vphantom{#1}}^{0}\kern-\scriptspace #1 }
\newcommand {\EvalOrderOne}[1] { \mathop{}\mathopen{\vphantom{#1}}^{1}\kern-\scriptspace #1 }
\newcommand {\EvalOrderTwo}[1] { \mathop{}\mathopen{\vphantom{#1}}^{2}\kern-\scriptspace #1 }
\newcommand {\KrawczykOrderTwo}[2] { \mathop{}\mathopen{\vphantom{#1}}^{2}\kern-\scriptspace K_{#1}(#2) } 

\newcommand {\CheckNoSolutionName}{\texttt{CertifyNoSolution}}
\newcommand {\CheckNoSolution}[2]{\CheckNoSolutionName\texttt{(}#1\texttt{,}#2\texttt{)}}

\newcommand {\CheckOneSolutionName}{\texttt{CertifyOneSolution}}
\newcommand {\CheckOneSolution}[2]{\CheckOneSolutionName\texttt{(}#1\texttt{,}#2\texttt{)}}

\newcommand {\IsSolInListName}{\texttt{IsSolInList}}
\newcommand {\IsSolInList}[3]{\IsSolInListName\texttt{(}#1\texttt{,}#2\texttt{,}#3\texttt{)}}

\newcommand {\CheckPrecUDName}{\texttt{CheckCond12}}
\newcommand {\CheckPrecUD}[4]{\CheckPrecUDName\texttt{(}#1\texttt{,}#2\texttt{,}#3\texttt{,}#4\texttt{)}}

\newcommand {\CheckPrecTName}{\texttt{CheckCond3}}
\newcommand {\CheckPrecT}[2]{\CheckPrecTName\texttt{(}#1\texttt{,}#2\texttt{)}}

\newcommand {\CheckPrecName}{\texttt{CheckPrec}}
\newcommand {\CheckPrec}[6]{\CheckPrecName\texttt{(}#1\texttt{,}#2\texttt{,}#3\texttt{,}#4\texttt{,}#5\texttt{,}#6\texttt{)}}

\newcommand {\SolveFixedPrecName}{\texttt{SolveWithFixedPrec}}
\newcommand {\SolveFixedPrec}[5]{\SolveFixedPrecName\texttt{(}#1\texttt{,}#2\texttt{,}#3\texttt{,}#4\texttt{,}#5\texttt{)}}

\newcommand {\BisectName}{\texttt{Bisect}}
\newcommand {\Bisect}[3]{\BisectName\texttt{(}#1\texttt{,}#2\texttt{,}#3\texttt{)}}

\RRetitle{A Subdivision Solver for Systems of Large Dense Polynomials}
\RRtitle{Un solveur par subdivision pour des syst\`emes de polyn\^omes denses de haut degr\'e et grande bit-size}
\RRauthor{
R\'{e}mi Imbach\thanks[VEGAS]{Loria Laboratory, INRIA Nancy Grand Est}
}

\RRdate{March 2016}
\RRversion{2}
\RRdater{October 2016}

\RRequipe{VEGAS}
\RCNancy

\RRabstract{We describe here the package {\tt subdivision\_solver} for the mathematical software {\tt SageMath}. 
It provides a solver on real numbers for square systems of large dense polynomials. 
By large polynomials we mean multivariate polynomials with large degrees, which coefficients have large bit-size. 
While staying robust, symbolic approaches to solve systems of polynomials see their performances dramatically affected by high degree and bit-size of input polynomials.
Available numeric approaches suffer 
from the cost of the evaluation of large polynomials and their derivatives.
Our solver is based on interval analysis and bisections of an initial compact domain of $\R^n$ where solutions are sought. 
Evaluations on intervals with Horner scheme is performed by the package {\tt fast\_polynomial} for {\tt SageMath}.
The non-existence of a solution within a box is certified by an evaluation scheme that uses a Taylor expansion at order 2, and existence and uniqueness of a solution within a box is certified with krawczyk operator.
The precision of the working arithmetic is adapted on the fly during the subdivision process and we present a new heuristic criterion to decide if the arithmetic precision has to be increased. }

\RRresume{
Le pr\'esent rapport d\'ecrit le package {\tt subdivision\_solver} pour {\tt SageMath}, d\'edi\'e \`a la r\'esolution r\'eelle de syst\`emes bien pos\'es (\emph{i.e.} avec autant d'\'equations que d'inconnues) dont les \'equations sont des polyn\^omes denses de haut degr\'es et grande bit-size.
Les approches de r\'esolution symboliques et num\'eriques voient leur performances tr\`es affect\'ees par l'augmentation du degr\'e et de la bit-size des polyn\^omes. Les premi\`eres en raison du calcul exact sur des grands nombres, les secondes \`a cause du co\^ut des \'evaluations des fonctions et de leur d\'eriv\'ees.
Le solveur d\'ecrit ici est bas\'e sur l'arithm\'etique par intervalles et s'appuie sur le package {\tt fast\_polynomial} pour {\tt SageMath} qui permet l'\'evaluation rapide de polyn\^omes sur des intervalles gr\^ace entre autres au sch\'ema de Horner.
Les solutions sont cherch\'ees dans un domaine initial compact de $\R^n$ qui est subdivis\'e jusqu'\`a pouvoir soit prouver l'absence de solutions gr\^ace \`a un d\'eveloppement de Taylor \`a l'ordre 2 soit prouver l'existence et l'unicit\'e d'une solution en utilisant l'op\'erateur de Krawczyk.
La pr\'ecision arithm\'etique avec laquelle les calculs sont r\'ealis\'es est adapt\'ee pendant la r\'esolution. On pr\'esente un nouveau crit\`ere, utilis\'e comme heuristique, pour d\'ecider s'il est n\'ecessaire d'accro\^itre la pr\'ecision.  
}

\RRkeyword{Interval Arithmetic, Subdivision, Adaptive Multi-Precision, Real Solutions, Large Dense Polynomials }
\RRmotcle{Arithm\'etique par intervalles, subdivision, pr\'ecision multiple adaptative, solutions r\'eelles, polyn\^omes denses de haut degr\'e et grande bit-size }

\begin{document}

 \RTNo{476}
 \makeRT
 
 \section{Introduction}
 
 \texttt{subdivision\_solver} is a solver for square systems of polynomial equations using exhaustive search in an initial bounded real domain given as a box (\emph{i.e.} a vector of intervals).
 It is specifically designed to handle systems of large dense polynomials and uses adaptive multi-precision arithmetic to stay robust to hard cases.  
 \texttt{subdivision\_solver} is proposed as a package for the mathematical software {\tt SageMath}\footnote{\url{http://www.sagemath.org}}. 
 It is based on the package {\tt fast\_polynomial} (see \cite{moroz:hal-00846961}) for {\tt SageMath} that provides fast evaluation of polynomials on intervals and uses Horner forms.  
 
 The strategies that underlies the design of \texttt{subdivision\_solver} are:
 \begin{itemize}
  \item using as much as possible double precision interval arithmetic: in this case interval computations are supported by the {\tt boost}\footnote{\url{http://www.boost.org/doc/libs/1_60_0/libs/numeric/interval/doc/interval.htm}} interval library that is efficient. When the arithmetic precision is increased, interval computations are supported by the Multi-Precision Floating-point Interval library {\tt mpfi} (see \cite{revol2005motivations}) that allows arbitrary arithmetic precision.
  \item using partial derivatives at second order to obtain sharp interval evaluations; partial derivatives are symbolically computed at the initialization of the solving.
  \item minimizing evaluations of polynomials and their derivatives that are costly.
 \end{itemize}

 The rest of this report is organized as follows. Sec.~\ref{sec_tools} gives some basics on interval analysis and describes the main tools the solver uses.
 Sec.~\ref{sec_precision} presents three criteria that are used to decide to increase arithmetic precision. The two first come from the state of the art, the third is new.
 Sec.~\ref{section_algorithm} describes the algorithms that we implemented.
 Sec.~\ref{sec_results} proposes some numerical results.
 The rest of the introduction overviews installation and usage of the package. 
 
 \subsection*{A short user manual}
 \texttt{subdivision\_solver} is a package for {\tt SageMath}, and works with versions newer than {\tt 7.0}.
 Its sole dependency is the package {\tt fast\_polynomial}\footnote{\url{http://www.loria.fr/~moroz/software.html}} for {\tt SageMath} that provides fast evaluation of polynomials on intervals and uses Horner forms.
 After it has been download, move to the directory where it is and install it with {\tt sage fast\_polynomial-0.9.4.spkg}.
 Then download \texttt{subdivision\_solver}\footnote{\url{http://www.loria.fr/~rimbach/}} and install it with the command {\tt sage subdivision\_solver-0.0.1.spkg}. Thats it. 
 
 We give here an example of use of {\tt subdivision\_solver} on randomly generated polynomials.
 
 \begin{verbatim}
  from subdivision_solver import subdivision_solver
  Rr.<x1,x2> =ZZ[] 
                             #pols of deg 100 with 1000 monomials
  p1 = Rr.random_element(100, 1000)
  p2 = Rr.random_element(100, 1000)
                             #set arithmetic precision 
                             #53 is the number of bits of the mantissa
  RIF = RealIntervalField(53) 
  X0 = [ [RIF(-1,1)], [RIF(-1,1)] ] 
                             #initialise the solver
  test = subdivision_solver([p1,p2],[x1,x2]) 
                             #search solutions in X0
                             #do not explore boxes of width smaller than 1e-6
                             #do not increase arithmetic precision above 113
                             #print stats in color
  status = test.solve(X0,1e-6,113,'stats_color')
                             #get solutions in sage lists
  sols = test.getSolutions()
                             #get isolating boxes with width smaller that 1e-10
  sols = test.getSolutions(1e-10)
                             #get boxes that may contain solutions
  unds = test.getUndetermined()
                             #get online help (only in console mode)
  subdivision_solver?
 \end{verbatim}
 
 Above instructions can be either copy-pasted in a {\tt sage} console to be executed, or copy pasted in a file {\tt myfile.sage} and executed with the command {\tt sage myfile.sage}.
 
 The initializer of the class {\tt subdivision\_solver} takes as arguments a list of polynomial and the list of variables of the polynomial ring.
 
 The method {\tt solve} takes as first argument the domain where the real solutions are sought, written as a column vector of {\tt RealIntervalField} elements. Notice that the arithmetic precision of the first element of the domain determines the initial precision.
 The arithmetic precision is at least 53. 
 When it is 53, bounds and intervals are double precision floating points, and operations on intervals are directly transposed into operations on {\tt double} with {\tt boost}.
 
 To solve the system given as input, the initial domain will be subdivided and/or contracted into sub-domains; 
 the second argument of the method solve is the minimum width of sub-domains to be explored during the resolution.
 It can be zero, and in this case the process is not guaranteed to terminate, in particular if the system admits a root with multiplicity.
 
 When solving the system given in input, it can arise that arithmetic precision is not sufficient when, for instance, subdividing a box does not help to decide where is a solution due to accumulation of errors in interval computations.
 Such cases are detected, and the arithmetic precision is increased to face it.
 The third argument of {\tt solve} is an upper bound of the arithmetic precision used to compute with intervals.
 Precision can be arbitrary large. {\tt mpfi} is used when the precision is larger than 53.
 
 The last argument of the method {\tt solve} is a string determining whether statistics of the process are printed or not. Its value can be in $\{$ {\tt 'silent'}, {\tt 'stats'}, {\tt 'stats\_color'} $\}$.
 
 The output of {\tt solve} is an integer with the following meaning:
 \begin{itemize}
  \item [$0$] Each solution lying in the interior of the initial domain has bean isolated in a box;
  isolating boxes of solutions are obtained with {\tt getSolutions}.
  \item [$1$] Only a subset of solutions have been isolated; boxes that may contain solutions are obtained with {\tt getUndetermined} but exploring these boxes will require a larger arithmetic precision than initially allowed.
  \item [$2$] Only a subset of solutions have been isolated; boxes of width smaller than allowed and/or requiring larger arithmetic precision where encountered.
 \end{itemize}

 Finally, giving a real number {\tt r} as argument to the method {\tt getSolutions} makes width of boxes isolating solutions that are returned smaller than {\tt r} if maximum precision allows it. Otherwise isolating boxes will have width as small as possible.
 
 \section{Some tools of Interval Analysis}
 \label{sec_tools}
 \subsection{Notations}
 Let $\IR$ be the set of non-empty closed intervals of $\R$. We will denote an element of $\IR$ with a lowercase bold letter as $\mInt{x}$.
 If $\mInt{x}\in\IR$, we will denote by $\borneinf{\mInt{x}}$ (resp. $\bornesup{\mInt{x}}$) its lower (resp. upper) bound,
 by $\width{\mInt{x}}$ its width defined as $\bornesup{\mInt{x}} - \borneinf{\mInt{x}}$ and 
 by $\middl{\mInt{x}}$ its center (that is $\borneinf{\mInt{x}} + \frac{\width{\mInt{x}}}{2}$).
 The interior $]\borneinf{\mInt{x}},\bornesup{\mInt{x}}[$ of $\mInt{x}$ will be denoted by $\interior{\mInt{x}}$ and its boundary $\mInt{x}\setminus\interior{\mInt{x}}$ by $\border{\mInt{x}}$.
 
 Let $m\in\N^+_*$ and consider $\IR^m$. We call \emph{box} an element of $\IR^m$ and we will denote it by an uppercase bold letter as $\mBox{X}$.
 Let $\mBox{X}=(\mInt{x}_1,\ldots,\mInt{x}_m)$ be an element of $\IR^m$, we will denote by $\middl{\mBox{X}}$ its center $(\middl{\mInt{x}_1},\ldots,\middl{\mInt{x}_m})$ and by $\width{\mBox{X}}$ its width defined as $\width{\mBox{X}}=max_{1\leq i \leq m}\width{\mInt{x}_i}$.
 We note $(\interior{\mInt{x}_1},\ldots,\interior{\mInt{x}_m})$ the interior of $\mBox{X}$ and $\border{\mBox{X}}$ its boundary $\mBox{X}\setminus\interior{\mBox{X}}$.
 
 Usual arithmetic operators can be extended to intervals and boxes.
 
 \subsection{Evaluating Functions on Boxes}
 
 Consider the polynomial function $f:\R^m\rightarrow \R$ and let $\mBox{X}\in\IR^m$. We will denote by $f(\mBox{X})$ the set $\{ f(X) | X\in\mBox{X}\}$.
 The following definitions are picked up from \cite{stahl1995}.
 
 We will call \emph{interval extension} of $f$ a function 
 $\intExt{f}:\IR^m\rightarrow \IR$ s.t. 
 \begin{equation}
 \label{eq:ext1}
  \mBox{X}\in\IR^m\Rightarrow f(\mBox{X})\subseteq\intExt{f}(X)
 \end{equation}
 and 
 \begin{equation}
 \label{eq:ext2}
  X\in\R^m\Rightarrow f(X)=\intExt{f}(X).
 \end{equation}
 We say that an interval extension $\intExt{f}$ is \emph{inclusion monotonic} if 
 \begin{equation}
  \label{eq:ext3}
  \mBox{X}\subseteq\mBox{Y}\Rightarrow \intExt{f}(\mBox{X})\subseteq\intExt{f}(\mBox{Y}).
 \end{equation}
 For a polynomial $f$, a natural interval extension of $f$ that is inclusion monotonic is obtained by combining interval operators that intervene in the Horner form of $f$ (see \cite{stahl1995}). We will denote it $\EvalOrderZero{f}$, and $\EvalOrderZero{f}(\mBox{X})$ will be the value of the evaluation of $\EvalOrderZero{f}$ on $\mBox{X}\in\IR^m$.
 
 An other inclusion monotonic interval extension (see \cite{stahl1995}) that is classically used is the mean value form $\EvalOrderOne{f}$ defined as 
 \begin{equation}
 \label{eq:EvalOrder1}
  \EvalOrderOne{f}(\mBox{X})= f(\mBox{P}) + \intExt{J_f}(\mBox{X})(\mBox{X}-\mBox{P})
 \end{equation}
 where $\mBox{P}$ is the center of $\mBox{X}$ represented as an interval and $\intExt{J_f}$ an inclusion monotonic interval extension of the Jacobian matrix $J_f$ of $f$ (\emph{i.e.} the $m$ components of $\intExt{J_f}$ are intervals extensions of the $m$ partial derivatives of $f$). $\EvalOrderZero{J_f}$, which components are
 $\EvalOrderZero{\frac{\partial f}{x_1}},\ldots,\EvalOrderZero{\frac{\partial f}{x_m}}$,
 is classically used for $\intExt{J_f}$.  
 
 Here we will use the interval extension obtained by considering Taylor expansion at order 2 around the center $\mBox{P}$ of $\mBox{X}$, defined as:
 \begin{equation}
 \label{eq:EvalOrder2}
  \EvalOrderTwo{f}(\mBox{X})= f(\mBox{P}) + J_f(\mBox{P})(\mBox{X}-\mBox{P}) + \frac{1}{2}(\mBox{X}-\mBox{P})^t (\EvalOrderZero{H_f}(\mBox{X})) (\mBox{X}-\mBox{P})
 \end{equation}
 where 
 $\EvalOrderZero{H_f}$ is the natural interval extensions of 
 the Hessian matrix $H_f$ of $f$. 
 $\EvalOrderTwo{f}$ is shown in \cite{rall1983mean} to be an inclusion monotonic interval extension of $f$, and we will call it evaluation at order two of $f$.
 
 Above definitions are easily extended to functions $F:\R^m\rightarrow \R^m$.
 Notice that if $\intExt{F}$ is an interval extension of $F$, then we have
 \begin{equation}
  \label{eq:notvanishing}
  0\notin\intExt{F} (\mBox{X})\Rightarrow F \text{ does not vanish on } \mBox{X}.
 \end{equation}
%
%
 
 \subsection{The Krawczyk Operator}
 
 We consider now a function $F:\R^m\rightarrow \R^m$ defined as $F(X)=(f_1(X),\ldots,f_m(X))$ where $f_i:\R^m\rightarrow \R$ for $1\leq i \leq m$ and we recall the definition of the Krawczyk operator $K_F$ that is a classical tool in interval analysis (see for instance \cite[Def.~1.16]{kearfott96} or \cite[Theo.~8.2]{MKCbook09} or \cite[Sec.~7]{Rump1983} ).
 \begin{equation}
  \label{eq:krawczyk}
 K_F(\mBox{X}) = \mBox{P} - (J_F(\mBox{P}))^{-1} F(\mBox{P}) + ( I - (J_F(\mBox{P}))^{-1}\intExt{J_F}(\mBox{X}))(\mBox{X} -\mBox{P})
 \end{equation}
 where $\mBox{P}$ is the center of $\mBox{X}$ represented as an interval and 
 $\intExt{J_F}$ is the interval extensions of 
 the Jacobian matrix $J_F$ of $F$. $K_f$ has the following properties:
 \begin{itemize}
  \item[$(K_1)$] $K_F(\mBox{X})\subset\interior{\mBox{X}}\Rightarrow F(X)=0$ has one and only one solution in $\mBox{X}$ that is in $K_F(\mBox{X})$
  \item[$(K_2)$] $K_F(\mBox{X})\cap\mBox{X}=\emptyset\Rightarrow F(X)=0$ has no solution in $\mBox{X}$
  \item[$(K_3)$] if $\mBox{X}$ and $\mBox{Y}$ are s.t. $K_F(\mBox{X})\subset\interior{\mBox{X}}$ and $K_F(\mBox{Y})\subset\interior{\mBox{Y}}$ then $\mBox{X}\cap\mBox{Y}\neq\emptyset\Rightarrow \mBox{X} \text{ and } \mBox{Y}$ contain the same solution of $F(X)=0$.  
 \end{itemize}
\textbf{Proof of $(K_3)$:} because $K_F$ is 
inclusion monotonic
(\emph{i.e.} $\mBox{X}\subseteq\mBox{Y}\Rightarrow K_{F}(\mBox{X})\subseteq K_{F}(\mBox{Y})$), see \cite{1984Krawczyk}, one has 
 $K_{F}(\mBox{X}\cap\mBox{Y})\subset\interior{\mBox{X}\cap\mBox{Y}}$ and $\mBox{X}\cap\mBox{Y}$ contains a unique solution.
 $\square$

 When using a mean value extension of $J_F$ around the center $\mBox{P}$ of $\mBox{X}$ to compute $\intExt{J_F}(\mBox{X})$ 
 one can rewrite Eq.~\ref{eq:krawczyk} as 
 \begin{equation}
 \label{eq:krawczykOrderTwo}
  \KrawczykOrderTwo{F}{\mBox{X}} = \mBox{P} - (J_F(\mBox{P}))^{-1}(F(\mBox{P}) + \mBox{H} )
 \end{equation}
 where $\mBox{H}$ is a box in $\IR^m$ which $i$-th component $\mInt{h_i}$ is:
 \begin{equation}
  \label{eq:hessian}
  \mInt{h_i} = (\mBox{X}-\mBox{P})^t (\EvalOrderZero{H_{f_i}}(\mBox{X})) (\mBox{X}-\mBox{P})
 \end{equation}
 where $\EvalOrderZero{H_{f_i}}$ is the natural interval extension of the Hessian matrix $H_{f_i}$ of $f_i$.
 We will call $\KrawczykOrderTwo{F}{\mBox{X}}$ the Krawczyk operator at order 2 since it uses second order partial derivatives of $F$; we assume here that it satisfies properties $(K_1),(K_2)$ and $(K_3)$.
 
 One can use $(K_2)$ and Eq.~\ref{eq:notvanishing} to justify Algo.~\ref{algo:checknosolution-algorithm} that certifies that a system $F=0$ with $F:\R^m\rightarrow \R^m$ has no solutions on a box $\mBox{X}$. 
 $(K_1)$ justifies Algo.~\ref{algo:checkonesolution-algorithm} that certifies that $F=0$ has a unique solution on $\mBox{X}$.
 Algo.~\ref{algo:checkunicity-algorithm} decides if the solution of $F=0$ contained in a box $\mBox{X}$ s.t. $K_F(\mBox{X})\subset\interior{\mBox{X}}$ is already contained in a box of a list $\mathcal{L}$ containing boxes $\mBox{Y}$ s.t. $K_F(\mBox{Y})\subset\interior{\mBox{Y}}$. It is correct from $(K_3)$.
 
 \begin{algorithm}[t]
    \caption{\CheckNoSolution{$F$}{$\mBox{X}$}}
\label{algo:checknosolution-algorithm}
\begin{algorithmic}[1]
 \Require{ A function $F:\R^m\rightarrow \R^m$ and a box $\mBox{X}$.}
 \Ensure{ {\bf true} or {\bf false}; if {\bf true} then $F=0$ has no solution in $\mBox{X}$. } 
 
 \If{ ($0\notin\EvalOrderZero{F}(\mBox{X})\cap\EvalOrderTwo{F}(\mBox{X})$) {\bf or} ($\KrawczykOrderTwo{F}{\mBox{X}}\cap\mBox{X}=\emptyset$)}
   \State \Return {\bf true}
 \EndIf 
 \State \Return {\bf false}
\end{algorithmic}
\end{algorithm}
\floatname{algorithm}{Algorithm}

\begin{algorithm}[t]
    \caption{\CheckOneSolution{$F$}{$\mBox{X}$}}
\label{algo:checkonesolution-algorithm}
\begin{algorithmic}[1]
 \Require{ A function $F:\R^m\rightarrow \R^m$ and a box $\mBox{X}$.}
 \Ensure{ {\bf true} or {\bf false}; if {\bf true} then $\mBox{X}$ contains a unique solution of $F=0$.} 
 \If{ $\KrawczykOrderTwo{F}{\mBox{X}}\subset\interior{\mBox{X}}$}
   \State \Return {\bf true}
 \EndIf
 \State \Return {\bf false}
\end{algorithmic}
\end{algorithm}
\floatname{algorithm}{Algorithm}

\begin{algorithm}[t]
    \caption{\IsSolInList{$F$}{$\mBox{X}$}{$\mathcal{L}$}}
\label{algo:checkunicity-algorithm}
\begin{algorithmic}[1]
 \Require{ A function $F:\R^m\rightarrow \R^m$, a box $\mBox{X}$ and a list of boxes $\mathcal{L}$ s.t. $K_F(\mBox{X})\subset\interior{\mBox{X}}$ and $\forall\mBox{Y}\in\mathcal{L}, K_F(\mBox{Y})\subset\interior{\mBox{Y}}$.}
 \Ensure{ {\bf true} or {\bf false}; if {\bf true} then the solution of $F=0$ contained in $\mBox{X}$ is in a box of $\mathcal{L}$.}
 \For{ $\mBox{Y}\in\mathcal{L}$ }
  \If{ $\mBox{X}\cap\mBox{Y}\neq\emptyset$ }
    \State \Return {\bf true}
  \EndIf
 \EndFor
 \State \Return {\bf false}
\end{algorithmic}
\end{algorithm}
\floatname{algorithm}{Algorithm}

 \section{Adapting Arithmetic Precision}
 \label{sec_precision}
 
 Intervals are usually represented by their two bounds represented by floating points, and interval operators by operations on floating points with appropriated rounding policy.
 In what follows, we will call \emph{precision} of the arithmetic the number of bits of the mantissa of floating points. 
 
 The branch and bound method we did implement uses recursive bisections of boxes until it is possible to certify either the absence or the existence and uniqueness of a solution in each box. Lack of accuracy can intervene during this process either when a box has a component that has roughly the smallest width allowed by the arithmetic precision (\emph{i.e.} the bounds are consecutive floating points), or when accumulated rounding affects the inclusion monotonicity of an interval extension.
 These cases have to be detected to avoid infinite computations. When they occur a simple strategy is to double the arithmetic precision and to continue the computations.
 
 We recall in \ref{subsec_revol} two criteria pickud up from \cite{revol03interval} that decide if the arithmetic precision is not sufficient, and we present in \ref{subsec_newcriterion} a criterion that is, as far as we know, new and is used as an heuristic in our solver.
 
 \subsection{Criteria of \cite{revol03interval}}
 \label{subsec_revol}
 
%
%
%
 
 \cite{revol03interval} presents a simple branch and bound algorithm using adaptive multi-precision to find zeros of an univariate real function $f$ with interval extension $\intExt{f}$, using interval newton method. 
 At each step, an interval $\mInt{x}$ is bisected in $\mInt{x}^1,\mInt{x}^2$ s.t. $\mInt{x}=\mInt{x}^1\cup\mInt{x}^2$, and the arithmetic precision is increased if one of the following conditions holds:
 \begin{itemize}
  \item $\width{\mInt{x}^1}\geq\width{\mInt{x}}$ or $\width{\mInt{x}^2}\geq\width{\mInt{x}}$,
  \item $\width{\intExt{f}(\mInt{x}^1)}\geq\width{\intExt{f}(\mInt{x})}$ or $\width{\intExt{f}(\mInt{x}^2)}\geq\width{\intExt{f}(\mInt{x})}$.
 \end{itemize}

 The first criterion is satisfied when $\mInt{x}$ has no point of its interior that can be represented as a floating point with actual precision. 
 When the second criterion is satisfied, Eq.~\ref{eq:ext3} is not true and $\intExt{f}$ is not an interval extension of $f$ due to numeric inaccuracy. One can alternatively test if $\intExt{f}(\mInt{x})\subseteq (\intExt{f}(\mInt{x}^1)\cup \intExt{f}(\mInt{x}^2))$.
 We adapt here these criteria to a multi-variate context with $F:\R^m\rightarrow \R^m$, $\intExt{F}:\IR^m\rightarrow \IR^m$ an interval extension of $F$, $\mBox{X}_1,\mBox{X}_2$ and $\mBox{X}$ elements of $\IR^m$ such that $\mBox{X}=\mBox{X}_1 \cup \mBox{X}_2$. In the branch and bound method presented in Sec.~\ref{section_algorithm}, the arithmetic precision is increased if one of the following conditions holds:
 \begin{itemize}
  \item[$(C_1)$] $\width{\mBox{X}_1}\geq\width{\mBox{X}}$ or $\width{\mBox{X}_2}\geq\width{\mBox{X}}$,
  \item[$(C_2)$] $\intExt{F}(\mBox{X})\subseteq(\intExt{F}(\mBox{X}_1)\cup\intExt{F}(\mBox{X}_2))$.
 \end{itemize}
 Algo.~\ref{algo:checkprecisionUD-algorithm} verifies these conditions and returns {\bf true} if $(C_1)$ or $(C_2)$ is satisfied.
 
 \begin{algorithm}[t]
    \caption{\CheckPrecUD{$F$ or $\mathcal{F}$}{$\mBox{X}$}{$\mBox{X}_1$}{$\mBox{X_2}$} }
\label{algo:checkprecisionUD-algorithm}
\begin{algorithmic}[1]
 \Require{ A function $F:\R^m\rightarrow \R^m$ or a list $\mathcal{F}$ of functions $f:\R^m\rightarrow\R$, three boxes $\mBox{X}$, $\mBox{X_1}$, $\mBox{X_2}$ .}
 \Ensure{ {\bf if} $(C_1)$ {\bf or} $(C_2)$ is satisfied {\bf then true}; {\bf false otherwise}.}
\end{algorithmic}
\end{algorithm}
\floatname{algorithm}{Algorithm}

 \subsection{A new criterion}
 \label{subsec_newcriterion}
 
 Consider Eqs. (\ref{eq:krawczyk}) and (\ref{eq:krawczykOrderTwo}).
 Whatever its order (we state it here for order two), one can rewrite Krawczyk operator as 
 \begin{equation}
  \KrawczykOrderTwo{F}{\mBox{X}} = - (\EvalOrderZero{J_F}(\mBox{P}))^{-1}(\EvalOrderZero{F}(\mBox{P})) + \mBox{P} + \text{something}
 \end{equation}
 to highlight that $F(\mBox{P})$ and $J_F(\mBox{P})$ are interval evaluations. Hence one has 
 \begin{equation}
  \width{\KrawczykOrderTwo{F}{\mBox{X}}} = \width{(\EvalOrderZero{J_F}(\mBox{P}))^{-1}(\EvalOrderZero{F}(\mBox{P}))} + \width{\text{something}}
 \end{equation}
 and a necessary condition for $\KrawczykOrderTwo{F}{\mBox{X}}$ to certify the existence of a unique solution in $\mBox{X}$ is $\width{(\EvalOrderZero{J_F}(\mBox{P}))^{-1}(\EvalOrderZero{F}(\mBox{P}))}<\width{\mBox{X}}$: otherwise one can not have $\KrawczykOrderTwo{F}{\mBox{X}}\subset \interior{\mBox{X}}$.
 Recall that $\mBox{P}$ is a point (\emph{i.e.} $\width{\mBox{P}}=0$) hence one should have, $\width{\EvalOrderZero{F}(\mBox{P})}=0$ and $\width{(\EvalOrderZero{J_F}(\mBox{P}))^{-1}(\EvalOrderZero{F}(\mBox{P}))}=0$.
 
 In the branch and bound method presented in Sec.~\ref{section_algorithm}, the arithmetic precision is increased if the following condition holds:
 \begin{itemize}
  \item[$(C_3)$] $\width{(\EvalOrderZero{J_F}(\mBox{P}))^{-1}(\EvalOrderZero{F}(\mBox{P}))}\geq\width{\mBox{X}} \text{ and } (\mBox{P} - (\EvalOrderZero{J_F}(\mBox{P}))^{-1}(\EvalOrderZero{F}(\mBox{P}))))\cap\mBox{X}\neq\emptyset$.
 \end{itemize}
 The condition $\width{(\EvalOrderZero{J_F}(\mBox{P}))^{-1}(\EvalOrderZero{F}(\mBox{P}))}\geq\width{\mBox{X}}$ can also be due to low values of partial derivatives of $F$ in $\mBox{P}$. 
 In that case, even if $\EvalOrderZero{F}(\mBox{P})$ has small width, its projection $\mBox{P} - (\EvalOrderZero{J_F}(\mBox{P}))^{-1}(\EvalOrderZero{F}(\mBox{P})))$ can have large width and can be located ``far away'' from $\mBox{X}$.
 The second part of the condition $(C_3)$ is used to filter the latter cases.
 Algo.~\ref{algo:checkprecisionT-algorithm} verifies $(C_3)$ and returns {\bf true} if it is satisfied.
 
 This condition is used as an heuristic and is checked only if a maximal arithmetic precision specified by the user has not been reached.
 
\begin{algorithm}[t]
    \caption{\CheckPrecT{$F$}{$\mBox{X}$} }
\label{algo:checkprecisionT-algorithm}
\begin{algorithmic}[1]
 \Require{ A function $F:\R^m\rightarrow \R^m$, a box $\mBox{X}$.}
 \Ensure{ {\bf if} $(C_3)$ is satisfied {\bf then true}; {\bf false otherwise}.}
\end{algorithmic}
\end{algorithm}
\floatname{algorithm}{Algorithm}
 
 \section{Algorithm}
 \label{section_algorithm}
 
 We describe in Algo.~\ref{algo:adaptive-algorithm} the adaptive multi-precision subdivision solver we implemented. Its input and output are fully specified in Sec.~\ref{subsec_InputOutput}.
 For a given initial box $\mBox{X}_0$ where solutions of $F(X)=0$ are sought, 
 a minimal width $\omega\in\R^+$ of boxes that can be explored, an initial arithmetic precision $p$ and a maximal precision $p_{max}$, it proceeds as follows.
 
 The simple branch and bound algorithm described in Algo.~\ref{algo:main-algorithm} is first applied to $\mBox{X}_0$ using $p$ as arithmetic precision.
 Its output are a list of boxes containing a unique solution of $F(X)=0$, 
 a list of boxes with width smaller than $\omega$ that can contain solutions and a list $\mathcal{L}_{prec}$ of boxes such that one of the conditions $(C_1),(C_2)$ or $(C_3)$ holds.
 
 The precision $p$ is then doubled and Algo.~\ref{algo:main-algorithm} is applied to boxes of $\mathcal{L}_{prec}$. This process is iterated until $p$ reaches $p_{max}$. When Algo.~\ref{algo:main-algorithm} is applied for the last time with $p=p_{max}$, the condition $(C_3)$ is not checked.
 
 Sec.~\ref{subsec_fixed} describes Algo.~\ref{algo:main-algorithm}.
 
 \begin{algorithm}[t]
    \caption{Subdivision Solving With Adaptive Arithmetic Precision}
\label{algo:adaptive-algorithm}
\begin{algorithmic}[1]
 \Require{ A function $F:\R^m\rightarrow \R^m$, a box $\mBox{X}_0$, a real number $\omega$, 
 an initial precision $p$, a maximal precision $p_{max}$ as described in Sec~\ref{subsec_InputOutput}.}
 \Ensure{ Two lists $\mathcal{L}_{sols}, \mathcal{L}_{comp}$ and an integer $r$ as described in Sec~\ref{subsec_InputOutput}.}
 
 \State Let $\mathcal{L}_{sols},\mathcal{L}_{sols}',\mathcal{L}_{comp},\mathcal{L}_{comp}'$ and $\mathcal{L}_{prec}$ be empty lists.
 \State Push back $(\mBox{X}_0,{\bf true})$ in $\mathcal{L}_{prec}$.
 \While { $\mathcal{L}_{prec}$ is not empty {\bf and} $p\leq p_{max}$ }
   \State $(\mathcal{L}_{sols}', \mathcal{L}_{comp}', \mathcal{L}_{prec})\leftarrow$ \SolveFixedPrec{$F$}{$\mathcal{L}_{prec}$}{$\omega$}{$p$}{$p_{max}$}
   \State Append boxes of $\mathcal{L}_{sols}'$ to $\mathcal{L}_{sols}$
   \State Append boxes of $\mathcal{L}_{comp}'$ to $\mathcal{L}_{comp}$
   \If{ $2p<p_{max}$ {\bf or} $p==p_{max}$ }
     \State Set $p$ to $2p$
   \Else
     \State Set $p$ to $p_{max}$
   \EndIf
 \EndWhile
 \State Determine $r$ as described in Sec~\ref{subsec_InputOutput}.
 \State \Return $\mathcal{L}_{sols}$, $\mathcal{L}_{comp}$, $r$
\end{algorithmic}
\end{algorithm}
\floatname{algorithm}{Algorithm}

 \subsection{Input and Output of Algo.~\ref{algo:adaptive-algorithm}}
 \label{subsec_InputOutput}
 
 Our multi-precision branch and bound method accepts mainly as input:
 \begin{itemize}
  \item a polynomial function $F(X)=(f_1(X),\ldots,f_m(X))$, where $X=(x_1,\ldots,x_m)$;
  \item a box $\mBox{X}_0$ where solutions of $F(X)=0$ are sought;
  \item a real number $\omega\in\R^+$ that is the minimum width of boxes to be explored during the subdivision process;
  \item an initial precision $p$ and a maximal precision $p_{max}$ of the floating arithmetic used to represent bounds of intervals, given as the number of bits of the mantissa.
 \end{itemize}
 The minimal width $\omega$ allows to avoid infinite computations arising in particular when $F(X)=0$ has non-regular solutions in $\mBox{X}_0$. If it is null, the termination of the process is not ensured.
 
 
 The output of our multi-precision branch and bound method consists in:
 \begin{itemize}
  \item a list of boxes $\mathcal{L}_{sols}$ containing solutions;
  \item a list of boxes $\mathcal{L}_{comp}$ where other solutions could lie;
  \item an integer $r$ specifying the status of the solving process.
 \end{itemize}

 Consider now the following properties:
 \begin{itemize}
  \item[$(P_1)$] if $\mBox{X}_s\in\mathcal{L}_{sols}$ then it exists a unique $X_s\in\mBox{X}_s\cap\interior{\mBox{X}_0}$ s.t. $F(X_s)=0$; 
  \item[$(P_2)$] if $X_s\in\interior{\mBox{X}_0}$ is s.t. $F(X_s)=0$ 
  then it exists at most one $\mBox{X}_s\in\mathcal{L}_{sols}$ s.t. $X_s\in\mBox{X}_s$; 
  \item[$(P_3)$] if $\mathcal{L}_{comp}$ is empty then for each $X_s\in\interior{\mBox{X}_0}$ s.t. $F(X_s)=0$ 
  it exists $\mBox{X}_s\in\mathcal{L}_{sols}$ s.t. $X_s\in\mBox{X}_s$; 
  furthermore there is no solution of $F=0$ on $\border{\mBox{X}_0}$.
 \end{itemize}

 Properties $(P_1)$ and $(P_2)$ are always satisfied when the process terminates.
 When, in addition, $(P_3)$ is satisfied, the value of $r$ is $0$.
 When $(P_3)$ is not satisfied because of a lack of arithmetic precision, the value of $r$ is $1$.
 Otherwise the value of $r$ is $2$.
 
 
 \subsection{Branch and Bound Method with Fixed Arithmetic Precision}
 \label{subsec_fixed}
 
 The branch and bound algorithm described in Algo.~\ref{algo:main-algorithm} can be seen as a 
 Depth First Search
 in a tree which root corresponds to an initial box $\mBox{X}_0$, and children of a node corresponding to a box $\mBox{X}$ correspond either to boxes obtained by bisecting $\mBox{X}$, \emph{i.e.} cutting it in two boxes with respect to one of its components, or to a box obtained by contracting $\mBox{X}$ around a solution.
 
 The goal of such an algorithm is to find sub-boxes $\mBox{X}$ of $\mBox{X}_0$ that contains a unique solution $X_{*}$ of $F=0$ s.t. $X_{*}\in\interior{\mBox{X}_0}$. Notice that the krawczyk test used in Algo.~\ref{algo:checkonesolution-algorithm} can only certify existence and uniqueness of a solution in $\interior{\mBox{X}}$, and solutions lying exactly on borders of sub-boxes of $\mBox{X}_0$ could be missed. $\epsilon$-inflation introduced for instance in \cite[Sec.~5.9]{stahl1995} can among other circumvent this pitfall. It consists in slightly enlarging a box $\mBox{X}$ obtaining $\mBox{X}_{\epsilon}$ s.t. $\mBox{X}\subset\mBox{X}_{\epsilon}$. Hence Algo.~\ref{algo:checkonesolution-algorithm} is applied to $\mBox{X}_{\epsilon}$ to certify existence and uniqueness of a solution in $\mBox{X}_{\epsilon}$. 
 This mechanism can however lead to find the same solutions in several neighbor boxes.
 $X_{*}\in\interior{\mBox{X}_0}$ is then verified when $\mBox{X}_{\epsilon}\subseteq\interior{\mBox{X}_0}$. 
 In order to optimize the number of evaluations of $F$ and its derivatives, the absence of solutions with Algo.~\ref{algo:checknosolution-algorithm} is also tested on $\mBox{X}_{\epsilon}$.
 
 A leaf of the tree corresponds 
 \begin{itemize}
  \item[$(n_1)$] either to a box $\mBox{X}$ s.t. $\width{\mBox{X}}\leq\omega$ that can possibly contain solutions but that are smallest than the minimum width $\omega$ given in input,
  \item[$(n_2)$] or to a box $\mBox{X}$ that is certified to contain no solution of $F=0$ with Algo.~\ref{algo:checknosolution-algorithm}, 
  \item[$(n_3)$] or to a box $\mBox{X}$ s.t. $\mBox{X}_{\epsilon}$ contains a unique solution $X_{*}$ of $F=0$ 
  and $\mBox{X}_{\epsilon}\subseteq\interior{\mBox{X}_0}$.
 \end{itemize}

 An inner node of the tree corresponds 
 \begin{itemize}
  \item[$(n_4)$] either to a box $\mBox{X}$ s.t. existence and uniqueness of a solution is certified in $\mBox{X}_{\epsilon}$, and s.t. $\mBox{X}_{\epsilon}\cap\border{\mBox{X}_{0}}\neq\emptyset$; this case is identified in Algo.~\ref{algo:main-algorithm} with the boolean variable {\tt inflateAndBisect};
  \item[$(n_5)$] either to a box $\mBox{X}$ where neither the absence nor the existence of a solution can be certified; such a node has two children corresponding to boxes obtained by bisecting $\mBox{X}$.
 \end{itemize}

 When a node corresponding to a box $\mBox{X}$ is visited, it is first checked with Algo.~\ref{algo:checkprecision-algorithm} that the actual arithmetic precision $p$ is sufficient. If it is not, $\mBox{X}$ is pushed in the list $\mathcal{L}_{prec}$; the subtree which root is the actual node will be explored with higher precision providing that $p<p_{max}$.
 Notice that the condition $(C_3)$ presented in Sec.~\ref{subsec_newcriterion} is checked in Algo.~\ref{algo:checkprecision-algorithm} only if $p<p_{max}$.
 
 If the arithmetic precision $p$ is sufficient, the type of the actual node ($(n_1)$ or $(n_2)$ or \ldots or $(n_5)$) is determined. 
 The actual node is of type $(n_3)$ if \CheckOneSolution{$F$}{$\mBox{X}_{\epsilon}$} returns {\bf true} (see Algo.~\ref{algo:checkonesolution-algorithm}) and $\mBox{X}_{\epsilon}\subseteq\interior{\mBox{X}_0}$.
 If the latter condition is not fulfilled the actual node is of type $(n_4)$ and the value of {\tt inflateAndBisect} is set to {\bf false}. The sole child of the actual node corresponds to the box $\KrawczykOrderTwo{F}{\mBox{X}_{\epsilon}}\subset \mBox{X}$.
 If $\mBox{X}_{\epsilon}\subseteq\interior{\mBox{X}_0}$, it is tested with Algo.~\ref{algo:checkunicity-algorithm} that the solution in $\mBox{X}_{\epsilon}$ has not already been found.
 Finally, when the actual node is of type $(n_5)$, $\mBox{X}$ is bisected.
 
 Literature proposes different strategies to choose a direction to cut a box $\mBox{X}$, and some of them are surveyed in \cite{just2014subdivision}. We did choose here to implement the maximum smear-diameter strategy (see \cite[Sec.~3.1.1, {\bf MaxSmearDiam}]{just2014subdivision}). In Algo.~\ref{algo:main-algorithm} the bisection is performed by the function \Bisect{$F$}{$\mBox{X}$}{$\omega$}, where the direction $i$ to cut $\mBox{X}=(\mInt{x}_1,\ldots,\mInt{x}_i,\ldots,\mInt{x}_m)$ is chosen so that $\width{\mInt{x}_i}>\omega$. The latter function returns a couple $(\mBox{X}_1,\mBox{X}_2)$ of boxes s.t. $\mBox{X}_1=(\mInt{x}_1,\ldots,[\borneinf{\mInt{x}_i},\middl{\mInt{x}_i}],\ldots,\mInt{x}_m)$ and $\mBox{X}_2=(\mInt{x}_1,\ldots,[\middl{\mInt{x}_i},\bornesup{\mInt{x}_i}],\ldots,\mInt{x}_m)$.    
 
 \begin{algorithm}[t]
    \caption{\SolveFixedPrec{$F$}{$\mathcal{L}_{work}$}{$\omega$}{$p$}{$p_{max}$}}
\label{algo:main-algorithm}
\begin{algorithmic}[1]
 \Require{ A function $F:\R^m\rightarrow \R^m$, a list $\mathcal{L}_{work}$ of boxes, a real number $\omega$, 
 a precision $p$ and a maximal precision $p_{max}$.}
 \Ensure{ Three lists $\mathcal{L}_{sols}, \mathcal{L}_{comp}$ and $\mathcal{L}_{prec}$ of boxes.}
 
 \While { $\mathcal{L}_{work}$ is not empty }
   \State Pop the front of $\mathcal{L}_{work}$ and store it in $(\mBox{X},{\tt inflateAndBisect})$.
   \If { \CheckPrec{$F$}{$\mBox{X}$}{$\omega$}{$p$}{$p_{max}$}{{\tt inflateAndBisect}} }
     \State Push back $\mBox{X}$ in $\mathcal{L}_{prec}$.
     \State {\bf continue}
   \EndIf
   \If { $\width{\mBox{X}} \leq \omega$ }
     \State Push back $\mBox{X}$ in $\mathcal{L}_{comp}$.
     \State {\bf continue}
   \EndIf
   \If{ {\tt inflateAndBisect} }
     \State Let $\mBox{X}_{\epsilon}$ be obtained by $\epsilon$-inflation of $\mBox{X}$.
   \Else
     \State Let $\mBox{X}_{\epsilon}$ be a copy of $\mBox{X}$.
   \EndIf
   \If{ \CheckNoSolution{$F$}{$\mBox{X}_{\epsilon}$} 
   } 
     \State {\bf continue}
   \EndIf
   \If{ \CheckOneSolution{$F$}{$\mBox{X}_{\epsilon}$} }
     \If{ 
     $\mBox{X}_{\epsilon}\cap\border{\mBox{X}_{0}}\neq\emptyset$ }
       \State Push back $(\KrawczykOrderTwo{F}{\mBox{X}_{\epsilon}},{\bf false})$ in $\mathcal{L}_{work}$
     \ElsIf{ $\mBox{X}_{\epsilon}\subseteq\interior{\mBox{X}_{0}}$ {\bf and not} \IsSolInList{$F$}{$\mBox{X}_{\epsilon}$}{$\mathcal{L}_{sols}$} }
       \State Push back $\mBox{X}_{\epsilon}$ in $\mathcal{L}_{sols}$
     \EndIf
   \Else
     \State $( \mBox{X}_1, \mBox{X}_2 ) =$ \Bisect{$F$}{$\mBox{X}$}{$\omega$}.
     \State Push $(\mBox{X}_1,{\bf true})$ and $(\mBox{X}_2,{\bf true})$ in the front of $\mathcal{L}_{work}$.
   \EndIf
 \EndWhile
 \State \Return $\mathcal{L}_{sols}$, $\mathcal{L}_{comp}$, $\mathcal{L}_{prec}$
\end{algorithmic}
\end{algorithm}
\floatname{algorithm}{Algorithm}

\begin{algorithm}[t]
    \caption{ \CheckPrec{$F$}{$\mBox{X}$}{$\omega$}{$p$}{$p_{max}$}{{\tt inflateAndBisect}} }
\label{algo:checkprecision-algorithm}
\begin{algorithmic}[1]
 \Require{ $F,\mBox{X},\omega,p,p_{max}$ as described in Sec.~\ref{subsec_InputOutput}, and a boolean variable {\tt inflateAndBisect}.}
 \Ensure{ {\bf true} if the arithmetic precision has to be increased to treat $\mBox{X}$; {\bf false} otherwise.}
 \If{ {\tt inflateAndBisect} }
   \State $( \mBox{X}^1, \mBox{X}^2 ) =$ \Bisect{$F$}{$\mBox{X}$}{$\omega$}.
   \State Let $\mBox{X}_{\epsilon}, \mBox{X}_{\epsilon}^1, \mBox{X}_{\epsilon}^2$ be obtained by $\epsilon$-inflation of $\mBox{X}, \mBox{X}^1, \mBox{X}^2$.
   \State \Return{ \CheckPrecUD{$F$}{$\mBox{X}_{\epsilon}$}{$\mBox{X}_{\epsilon}^1$}{$\mBox{X}_{\epsilon}^2$} {\bf or} ( \CheckPrecT{$F$}{$\mBox{X}_{\epsilon}$} {\bf and} $p<p_{max}$ ) } 

 \Else
   \State \Return{(\CheckPrecT{$F$}{$\mBox{X}$} {\bf and} $p<p_{max}$) 
   }
   
 \EndIf
\end{algorithmic}
\end{algorithm}
\floatname{algorithm}{Algorithm}
 
 \section{Benchmarks}
 \label{sec_results}
 
 We propose here some experimental results to justify the chosen strategy for certifying non-existence or existence and uniqueness of a solution in a box.
 We compare four possible strategies for functions \CheckNoSolutionName~ and \CheckOneSolutionName~:
 \begin{itemize}
  \item[(1)] Evaluation at order 2 and Krawczyk operator at order 2, as described in previous algorithms;
  \item[(2)] Evaluation at order 2 and Krawczyk operator as defined in Eq.~\ref{eq:krawczyk}: obtained by replacing $\KrawczykOrderTwo{F}{X}$ by $K_F(X)$ in previous algorithms;
  \item[(3)] Evaluation at order 1 and Krawczyk operator as defined in Eq.~\ref{eq:krawczyk}: obtained by replacing $\EvalOrderTwo{F}$ by $\EvalOrderOne{F}$ and $\KrawczykOrderTwo{F}{X}$ by $K_F(X)$ in previous algorithms;
  \item[(4)] Natural interval extension and Krawczyk operator as defined in Eq.~\ref{eq:krawczyk}: obtained by replacing $\EvalOrderTwo{F}$ by $\EvalOrderZero{F}$ and $\KrawczykOrderTwo{F}{X}$ by $K_F(X)$ in previous algorithms.
 \end{itemize}

 We did use these four strategies to solve dense, randomly generated, systems with $m$ polynomial equations of degree $d$ with integer coefficients of bit-size $8$, for $(m,d)\in\{ (2,64), (2,128), $ $(3,16),(3,32), (4,8), (5,4)  \}$.
 Table~\ref{tab:strategies} gives for each strategy and each couple $(m,d)$ the number $n$ of boxes that have been explored and the sequential time $t$ in seconds spent to solve the system.
 
 We note that using evaluation at order 2 (\emph{i.e.} strategies (1) and (2) versus (3) and (4)) brought an important gain both in terms of number of boxes explored and time, and higher $m$ and $d$ are, higher is this gain.
 The gain allowed by Krawczyk operator at order 2 (strategy (1) versus strategy (2)) is more perceptible for small values of $m$ and $d$. Notice that applying strategy (1) instead of strategy (2) does not induce additional evaluations since values of the Hessian matrix are computed for the evaluation at order 2. 
 
 \begin{table}
  \caption{ Comparaison of strategies (1),(2),(3) and (4). 
  $m$ is the number of equations and variables of the system, $d$ is the degree of its polynomial equations.
  $n$ is the number of boxes that have been explored, $t$ the sequential time in seconds spent for the process on a 
  Intel(R) Core i7-5600U CPU @ 2.60GHz .}
\begin{footnotesize}
\begin{center}
\begin{tabular}{r|cc|cc||cc|cc||cc||cc|}
$m$  & \multicolumn{4}{|c||}{2}                              & \multicolumn{4}{|c||}{3}                             & \multicolumn{2}{|c||}{4}& \multicolumn{2}{|c|}{5} \\\hline
$d$  & \multicolumn{2}{|c|  }{64}& \multicolumn{2}{|c||}{128}& \multicolumn{2}{|c| }{16}& \multicolumn{2}{|c||}{32} &\multicolumn{2}{|c||}{8} &\multicolumn{2}{|c| }{4} \\\hline
     &  $n$       &    $t$       &  $n$       &    $t$       &  $n$      &    $t$       &   $n$       &    $t$      &   $n$      &    $t$     &    $n$     &    $t$         \\\hline
(1)  &   855      &    0.33      &   1028     &    1.75      &    6650   &     2.65     &     18310   &    61.5     &    49647   &    17.0    &    104373  &    10.7     \\\hline
(2)  &   886      &    0.40      &   1053     &    2.01      &    6943   &     3.16     &     18881   &    70.9     &    52501   &    20.3    &    110229  &    12.6     \\\hline
(3)  &   1158     &    0.36      &   1594     &    2.24      &   14338   &     4.05     &     47703   &    107      &    158076  &    36.4    &    298727  &    21.9      \\\hline
(4)  &   1286     &    0.40      &   1916     &    2.66      &   23219   &     6.62     &     102539  &    230      &    363274  &    81.3    &    576107  &    39.6      \\\hline
\end{tabular}
\end{center}
\end{footnotesize}
\label{tab:strategies}
\end{table}
 

\end{document}